# Numerical Simulation of Jet Mode in Electrospraying of Newtonian and Viscoelastic Fluids

**Amirreza Panahi [a], Ahmad Reza Pishevar [a,1], Mohammad Reza Tavakoli[a]**

[a] *Department of Mechanical Engineering, Isfahan University of Technology, Isfahan 84156-8311, Iran*

## Abstract
The purpose of this article is to explore the role of viscoelastic properties of polymeric solutions on the jet mode in the electrospray process. In this research, several numerical simulations were performed to model the behavior of electrified Newtonian and viscoelastic jets. First, used for validating viscoelastic constitutive equations and their implementation, the benchmark problem of the sedimenting sphere is stabilized beyond the previously suggested threshold. Then, an electrified DI water jet was simulated, and the obtained jet profile was compared with the experimental data from previous publications. Finally, the proposed algorithm was used to simulate viscoelastic electrified jets, where the effect of the Weissenberg number (Wi) on the jet profile was examined. In agreement with the previously obtained experimental results, by increasing the solution concentration, the asymptotic profile of the jet is reached at a smaller length from the nozzle, while the final thickness of the jet is slightly reduced.

## I. Introduction

Electrospray process, a method of producing a continuous stream of monodisperse droplets in ambient air, has received increasing attention in the past years. The procedure of conducting an electrospray test includes applying a potential difference between a nozzle that is ejecting a continuous stream of droplets and a substrate positioned directly below the nozzle. Electrospray has been presented as one of the most beneficial methods for producing monodisperse droplets due to the flexibility it allows for droplet size and distribution. This method has several applications including thin-film deposition[1], polymer particle production[2], and nanoparticle preparation[3, 4].

The applied potential difference can produce different electrohydrodynamic (EHD) modes in the electrospraying of a fluid with specific physical properties. These modes are classified according to the shape and behavior of the fluid that is ejected from the nozzle. For instance, the EHD modes observed in Newtonian fluids have been categorized in several studies[5-7], and they include the dripping, microdripping, spindle, oscillating-jet, cone-jet, and precession modes. As for viscoelastic fluids, the observed EHD modes were identified in our previous study, and they comprise the dripping mode, beads-on-a-string structure, cone-jet mode, stick jet mode and unstable jet. The jet mode of viscoelastic fluids, consisting of both cone-jet and stick jet modes, has been simulated in this work, and its underlying physical mechanisms will be explored in detail in Sec. III.

[1] Author to whom correspondence should be addressed: apishe@cc.iut.ac.ir

The observed EHD modes have led many researchers to study and model the dynamic response of fluids to electric fields applied to them. Initially, most studies were focused on perfect dielectric fluid or perfect conductive fluid models. It was not until the pioneering work of Taylor[8] and Taylor and Melcher[9] that the leaky dielectric model became known as a way to model fluid deformation through the accumulation of electric charges on the two-phase flow interface. Ever since the introduction of this model, numerical and theoretical studies pertinent to this model have become ubiquitous in the literature. Saville[10] summarized the main concepts and equations for this model in a 1997 review article. Moreover, in several previous studies, this model was used to simulate the deformation of a drop within a specified electric field strength[11-14] and the well-known oblate and prolate deformations and conditions under which these deformations were reached were thoroughly discussed. Due to its ability to exert tangential force on the fluid interface, the leaky dielectric model has been utilized for cone-jet simulation[15-18] and electrically controlled droplet generation[19].

The Weissenberg number (Wi), a dimensionless value that is regularly used in problems of viscoelastic fluids, is inherently large in electrospray problems. As reported in previous numerical studies, measures have to be taken to deal with the high Wi problem (HWNP). Different techniques have been proposed to stabilize numerical solutions in cases of high Wi values. The inconsistent streamline upwinding (SU) method, a special case of the Petrov–Galerkin formulation, can be used to stabilize numerical solutions for convection-dominant problems. Other stabilization methods, such as adding weak-form stabilization terms, have been proposed by Behr et al.[20] and Coronado et al.[21] for the Oldroyd-B[22] model. Additionally, the SU method, together with the log-conformation method (LCM), can be used to stabilize the numerical solution of the Oldroyd-B model for viscoelastic fluid flow at high Wis.

The LCM reformulation, initially proposed by Fattal and Kupferman[23, 24], solves the logarithm of the conformation tensor. In this way, the positive definiteness of the conformation tensor is preserved, and the extensional components of the deformation field behave additively. The LCM reformulation has been used to solve several sophisticated problems of viscoelastic fluid flow, including lid-driven cavity stokes flow[25, 26], flow past a confined circular cylinder[27-29], flow past a sphere in a cylindrical tube[30, 31], abrupt contraction[32, 33], viscoelastic flow in a curvilinear microchannel[34] and viscoelastic extrudate swell[35, 36].

The methods introduced in the literature to properly model the moving interface in two-phase flows include the volume-of-fraction method, the level-set method, and the phase-field method. These methods are widely used in both Newtonian and viscoelastic two-phase flows, however, the phase-field method is used in our simulations due to the improved numerical convergence.

The phase-field method has been the main subject of several studies, where it has been introduced as a versatile tool in multi-phase flow modeling. Two main types of this model include the Allen–Cahn and Cahn–Hilliard equations. The governing equations for both types have been delineated and investigated in the literature[37, 38]; however, this article largely focuses on the Cahn–Hilliard equation, especially in conjunction with the Navier–Stokes equation[39-41]. On top of that, the Cahn–Hilliard equation has been successfully coupled with different viscoelastic fluid models in previous works[42-44] that investigate the applicability of the phase-field model to solve multi-phase non-Newtonian fluid flow.

Many studies have investigated the role of viscoelasticity on the flow behavior theoretically and numerically. For example, the atomization mechanism of a charged viscoelastic liquid sheet was examined in one work by solving viscoelastic constitutive equations in a perturbed state[45]. The electrospinning of polyisobutylene-based solutions was investigated theoretically by Carroll and Joo[46] and numerically using the FENE-P model by Zhmayev et al.[47]. Additionally, the atomization of polymer solutions was numerically modeled by Qian et al.[48]. Li et al.[49] used Oldroyd-B and leaky dielectric models to investigate viscoelastic jet axisymmetric and non-axisymmetric instabilities. The beads-on-a-string structure, which is seen in viscoelastic fluids due to the delayed breakup process, was numerically modeled by Turkoz et al.[50].

The main goal of this research is to explore the role of viscoelasticity on jet mode. Therefore, the process is investigated numerically by solving the constitutive equations for viscoelastic electrified jets. In our case, due to the small characteristic length of the problem, the Wi is rather large. As a result, the HWNP, a complication encountered in the solution of viscoelastic constitutive equations, is addressed and the implementation of the LCM reformulation is clearly described to rectify the problem. To validate the overall results of the current work, the jet profiles obtained through numerical simulations are compared with their corresponding experimental data, and a good agreement is seen between the two data sets.

The rest of this paper is organized as follows: the problem is formulated in Sec. II. In Sec. III, the results of simulations are discussed thoroughly. Finally, the paper is concluded in Sec. IV.

## II. Governing Equations

Here, we assume that the system under consideration consists of two immiscible, incompressible fluids. One, the polymeric solution, behaves as a viscoelastic fluid, while the other one, the air phase, behaves as a Newtonian fluid. In the following subsections, the equations used to model the current system are given step by step descriptions. Subsequently, the aforementioned equations are coupled in fluid flow equations. The governing equations in this section are delineated in axisymmetric coordinates, with the z-axis considered the symmetry axis.

### A. Viscoelastic Constitutive Equations

The flow behavior of viscoelastic fluid can be explained by generalized Newtonian, linear viscoelastic or non-linear viscoelastic models[51]. The Oldroyd-B model, a simple non-linear viscoelastic model, is chosen as our governing equation to model the flow behavior of the viscoelastic fluid. Assuming a constant polymeric viscosity (no shear thinning), which makes it a perfect fit for modeling Boger fluids, this model is written as follows:

$$\tau + \lambda_1 \tau^{\nabla} = 2\eta_0(\dot{\gamma} + \lambda_2 \dot{\gamma}^{\nabla}) \quad (1)$$

where $\lambda_1$, $\lambda_2$, $\eta_0$ and $\dot{\gamma}$ denote relaxation time, retardation time, total viscosity, and deformation rate, respectively. The total viscosity is defined as the sum of polymeric and solvent contributions to viscosity, and the deformation rate tensor is defined by the velocity gradient:

$$\eta_0 = \eta_p + \eta_s \quad (2)$$

$$\dot{\gamma} = 0.5(\nabla u + (\nabla u)^T) \quad (3)$$

The $\tau^{\nabla}$ variable is the upper convected Maxwell derivative of stress and is defined as follows:

$$\tau^{\nabla} = \frac{\partial \tau}{\partial t} + (u.\nabla)\tau - (\nabla u)^T.\tau - \tau.(\nabla u) \quad (4)$$

It should be noted that the definition of the upper convected Maxwell derivative for the deformation rate is identical to Eq. (4). By splitting the total stress and relating relaxation and retardation times, we produce the following[51]:

$$\tau = \tau_p + \tau_s \tag{5}$$

$$\lambda_2 = \left(\frac{\eta_s}{\eta_p + \eta_s}\right)\lambda_1 \tag{6}$$

If Eqs. (5) and (6) are substituted into Eq. (1), the following relationships are obtained:

$$\tau_s = 2\eta_s \dot{\gamma} \tag{7}$$

$$\tau_p + \lambda_1 \overset{\nabla}{\tau_p} = 2\eta_p \dot{\gamma} \tag{8}$$

Eqs. (7) and (8) show that Newtonian and polymeric stress equations can be solved and added independently to the Navier–Stokes equation. From this point forward, the index of $\lambda_1$ is omitted for simplicity. If a variable change is used for polymeric stress, then Eq. (9) obtains:

$$\tau_p = \frac{\eta_p}{\lambda}(\sigma - I) \tag{9}$$

If Eq. (9) is substituted into Eq. (8), we obtain the constitutive equation of conformation tensor, as below:

$$\frac{\partial \sigma}{\partial t} + (u.\nabla)\sigma - \nabla u^T.\sigma - \sigma.\nabla u = \frac{1}{\lambda}(I - \sigma) \tag{10}$$

Then, the divergence of polymeric stress is added to the Navier–Stokes equation as a volume force:

$$F_p = \nabla.\tau_p = \nabla.\left(\frac{\eta_p}{\lambda}\sigma\right) \tag{11}$$

Let us define Wi as follows:

$$Wi = \lambda \frac{U}{l} \tag{12}$$

where $l$ is the characteristic length of the problem. Previous studies have found that, regardless of the numerical scheme used for the discretization of Eq. (10), the solution will fail for relatively high Wis[52]. This problem has been the main obstacle to numerical rheology in recent decades. The stability threshold for our numerical simulations is increased by the implementation of LCM reformulation. This method begins with a unique decomposition of the velocity gradient transpose:

$$\nabla u^T = \Omega + \mathrm{B} + \mathrm{N}\sigma^{-1} \tag{13}$$

where N and Ω are anti-symmetric tensors, and B is a symmetrical, traceless tensor. For the sake of clarity, we define the velocity gradient tensor in axisymmetric coordinates as follows:

$$\nabla u = \begin{pmatrix} \frac{\partial u}{\partial r} & 0 & \frac{\partial w}{\partial r} \\ 0 & \frac{u}{r} & 0 \\ \frac{\partial u}{\partial z} & 0 & \frac{\partial w}{\partial z} \end{pmatrix} \tag{14}$$

Following Eq. (14), the general form of the viscoelastic stress tensor can be written as follows:

$$\tau_p = \begin{pmatrix} \tau_{11} & 0 & \tau_{13} \\ 0 & \tau_{22} & 0 \\ \tau_{13} & 0 & \tau_{33} \end{pmatrix} \tag{15}$$

The general form of conformation and log-conformation tensors is similarly defined. Furthermore, a symmetrical positive definite tensor, the conformation tensor can be decomposed as follows:

$$\sigma = R \Lambda R^T \tag{16}$$

where $R$ is an orthogonal tensor made by the eigenvectors of $\sigma$, and ʌ is a diagonal tensor made by eigenvalues of $\sigma$. Next, N, Ω, and B are decomposed by the tensor $R$ and its transpose:

$$N = R \begin{pmatrix} 0 & n_{12} & n_{13} \\ -n_{12} & 0 & n_{23} \\ -n_{13} & -n_{23} & 0 \end{pmatrix} R^T \tag{17}$$

$$\Omega = R \begin{pmatrix} 0 & \omega_{12} & \omega_{13} \\ -\omega_{12} & 0 & \omega_{23} \\ -\omega_{13} & -\omega_{23} & 0 \end{pmatrix} R^T \tag{18}$$

$$B = R \begin{pmatrix} b_{11} & 0 & 0 \\ 0 & b_{22} & 0 \\ 0 & 0 & b_{33} \end{pmatrix} R^T \tag{19}$$

Tensor $Q$ is defined as follows:

$$R^T(\nabla u^T)R = Q = \begin{pmatrix} q_{11} & q_{12} & q_{13} \\ q_{21} & q_{22} & q_{23} \\ q_{31} & q_{32} & q_{33} \end{pmatrix} \tag{20}$$

Eqs. (17)–(20) are used to decompose the velocity gradient. For more details, readers are referred to Appendix A. If we substitute Eq. (13) into Eq. (10) and simplify the result, following Fattal and

Kupferman[23], the log-conformation constitutive equation is obtained:

$$\frac{\partial\psi}{\partial t} + (u.\nabla)\psi - (\Omega\psi - \psi\Omega) - 2B = \frac{1}{\lambda}\left(e^{-\psi} - I\right) \quad (21)$$

where $\psi$ is the log-conformation tensor. The conformation and log-conformation tensors are related by the eigenvectors of the conformation tensor:

$$\psi = R\log(\Lambda^{\sigma})R^{T} \quad (22)$$

$$\sigma = Rexp\left(\Lambda^{\psi}\right)R^{T} \quad (23)$$

where $\Lambda^{\sigma}$ and $\Lambda^{\psi}$ are tensors made by eigenvalues of conformation and log-conformation tensors, respectively. The relations used to acquire the eigenvalues and eigenvectors of the log-conformation tensor are shown in Appendix A.

**B. Two-phase Flow Equations**

To capture the interface, the phase-field method is adopted, and the surface tension force is applied to every node near the interface as the body force. First, the phase-field parameter is defined as follows:

$$\emptyset = \frac{m_1 - m_2}{m_1 + m_2} \quad (24)$$

where $m_1$ and $m_2$ are the masses of each phase. Alternatively, Eq. (24) can be interpreted as indicating the differences in concentration between two phases, where the concentration for each phase has a value between 0 and 1, and consequently, the phase-field parameter can vary between -1 and 1. The Helmholtz free energy for unit volume in a homogenous mixture is defined as follows[53]:

$$F(\emptyset) = \frac{1}{4}(\emptyset^2 - 1)^2 \quad (25)$$

Additionally, the chemical potential is defined as follows:

$$f(\emptyset) = F'(\emptyset) - \varepsilon^2\Delta\emptyset \quad (26)$$

where $\varepsilon$ is the interface thickness. The Cahn–Hilliard equation is a conserved form for the phase-field model, and it is written as follows:

$$\frac{\partial\emptyset}{\partial t} + u.\nabla\emptyset = \nabla.\left(M(\emptyset)\nabla f(\emptyset)\right) \quad (27)$$

where $M(\emptyset)$, denoting mobility, has been given various definitions in the literature. In the phase-field method, the surface tension force is obtained by the following[40]:

$$F_{ST} = \frac{\lambda}{\varepsilon^2} f(\emptyset)\nabla\emptyset \quad (28)$$

where $\lambda$ is the mixing energy density. Then, the volume fraction of each phase is defined by phase-field parameter, as follows:

$$Vf_1 = \frac{1-\emptyset}{2} \quad \& \quad Vf_2 = \frac{1+\emptyset}{2} \quad (29)$$

**C. Electrostatic Equations**

As previously noted, the leaky dielectric model is used to simulate the electric field. Neither perfectly dielectric nor perfectly conductive, poorly conductive fluids are typically modeled with the leaky dielectric model, where the effect of the accumulation of charge on an infinitely thin interface is taken into account. As shown by Saville[10], magnetic effects in EHD problems can be omitted because the characteristic time for magnetic phenomena is several orders of magnitude smaller than the characteristic time for electric phenomena. Therefore, it is only necessary to deal with general electrostatic equations:

$$\nabla \times E = 0 \quad (30)$$

$$E = -\nabla V \quad (31)$$

where $E$ is the electric field and $V$ is the electric potential. The space charge density is related to the electric field (or potential difference) through the following relations:

$$\nabla.(\varepsilon_0\varepsilon E) = -\varepsilon_0\nabla.(\varepsilon\nabla V) = \rho_e \quad (32)$$

where $\rho_e$, $\varepsilon_0$ and $\varepsilon$ are space charge density, the vacuum permittivity and relative permittivity, respectively. Furthermore, the following charge-conservation law should be satisfied at every node[11]:

$$\frac{\partial\rho_e}{\partial t} + \nabla.J = 0 \quad (33)$$

where $J$ is the current density, defined as:

$$J = \sigma E + \rho_e u \quad (34)$$

The first term in Eq. (34) represents ohmic charge conduction, and the second term represents charge convection by the velocity field. Before any further simplification, the charge relaxation time, viscous relaxation time, and capillary time scale are defined as follows:

$$\tau_c = \frac{\varepsilon_0 \varepsilon}{\sigma} \qquad (35)$$

$$\tau_\mu = \frac{\rho l^2}{\mu} \qquad (36)$$

$$\tau_{cap} = \sqrt{\frac{\rho l^3}{\gamma}} \qquad (37)$$

In a leaky dielectric system, electric charges are accumulated near the interface, and it is assumed that the thickness of the electric double layer is very small relative to the scale of the problem length. In this case, the diffusion of electric charges is neglected, so the space charge density is assumed to be zero and the effects of surface charges are considered to be a boundary condition[12, 13]. The electric relaxation time is small relative to the viscous time scale; consequently, the charge-conservation equation is simplified by omitting the convection term and its quasi-static form is considered[15]:

$$\nabla.(\sigma E) = 0 \qquad (38)$$

In the dripping mode, the capillary time scale has the same order of magnitude as the electric relaxation time, so Eq. (38) is only plausible for the simulation of the jet mode[15]. Finally, the Maxwell stress tensor is defined as follows:

$$\tau^e = \varepsilon_0 \varepsilon \left( \vec{E}\vec{E} - \frac{1}{2}\, \vec{E}.\vec{E} \right) \qquad (39)$$

The divergence of the Maxwell stress tensor yields the force exerted on the ejecting fluid by the electric field:

$$F^e = \nabla.\tau^e = -\frac{1}{2}\, \vec{E}.\vec{E}\, \nabla \varepsilon_0 \varepsilon + \rho_e \vec{E} \qquad (40)$$

The first term in Eq. (40), the dielectric force, represents the force exerted due to the polarization effects, and the second term, the Coulomb force, is the force applied on space charges.

**D. Fluid Flow Equations**

The fluid flow equations include incompressible continuity and the Navier–Stokes equation:

$$\nabla . u = 0 \qquad (41)$$

$$\rho \frac{\partial u}{\partial t} + \rho (u.\nabla) u = \nabla . \beta + \rho g + F_{ST} + F_e + F_p$$

(42)

where $F_p$, $F_{ST}$ and $F_e$ are previously defined by Eqs. (11), (28), and (40), respectively. Additionally, $\beta$ comprises the pressure and Newtonian viscous stress, as follows:

$$\beta = -pI + 2\eta_s \dot{\gamma} \qquad (43)$$

Inevitably, every physical property described in the above equations should be smeared across the interface through the volume fraction of fluids. For instance, the following relation is written for the relaxation time:

$$\lambda_r = \lambda_1 Vf_1 + \lambda_2 Vf_2 \qquad (44)$$

where $Vf$ is the volume fraction of corresponding fluids. The physical properties of viscoelastic fluids used in simulations are equal to the properties of polyacrylamide (PAA) solutions in three different concentrations including 50, 100 and 150 ppm. The physical properties of the PAA solutions, DI water and air utilized in the current calculations were measured in our previous work, where these properties are classified in two tables. Since negligible shear-thinning behavior was observed in small-amplitude oscillatory shear tests (the polymeric viscosity remained fairly constant for different shear rates), the Oldroyd-B model can predict viscoelastic behavior for PAA solutions with acceptable accuracy. It is worth noting that the polymeric viscosity and relaxation time are zero in the air; nevertheless, to avoid numerical complications, a very small value ($1e^{-14}$) is considered for the air relaxation time and polymeric viscosity.

### E. Dimensionless Numbers

In this subsection, the dimensionless numbers influencing the electrospray process are introduced. The Weber number, the ratio of inertia forces to surface tension forces, is defined as follows:

$$We = \frac{\rho u^2 l}{\gamma} \quad (45)$$

The electric capillary number is the ratio of the electric field forces to the surface tension forces and is given by:

$$Ca_E = \frac{\varepsilon_0 \varepsilon_r E^2 R_0}{\gamma} \quad (46)$$

where $R_0$ is half of the characteristic length, which in our case is the outer radius of the nozzle. For the calculation of the electric field strength, the relationship proposed by Jones and Thong[54] for the electric field at the tip of a positively charged cylinder with a semi-finite ground terminal positioned below the cylinder was used:

$$E = \frac{\sqrt{2}\emptyset_0}{R_0 \ln(4z_0/R_0)} \quad (47)$$

where $z_0$ is the distance between cylinder and ground terminal, and $\emptyset_0$ is the applied voltage. Additionally, $\varepsilon_r$ in Eq. (46) is the characteristic relative permittivity, which is derived using the Lorentz model for the interaction of electromagnetic waves in dielectric materials[55]:

$$\varepsilon_r = 1 + \frac{\sigma}{\varepsilon_0 \omega} \quad (48)$$

where $\omega$ is characteristic frequency:

$$\omega = \frac{c}{L} \quad (49)$$

In Eq. (49), $c$ is the speed of light in the air and $L$ is the distance between the center of the capillary tip and the inner edge of an annular disk taken as the substrate. These dimensionless numbers are utilized in Sec. III to categorize and classify the results.

## III. Results

The governing equations introduced in Section II are discretized with the Petrov–Galerkin finite-element method in axisymmetric coordinates. The coupling between the electric and viscoelastic stresses and flow equations is accomplished using an iterative segregated approach, and the non-linear system of equations is solved using the Newtonian method. For every iteration in a new timestep, first, the electric-potential equation is solved. Then, using the electric stresses obtained from the previous step and the viscoelastic stresses obtained from the previous iteration, the flow equations and the phase-field equation are solved to update the velocity components and the interface position. Finally, the LCM equations are solved using the updated velocity components and the interface position. Iterations continue until a simultaneous convergence is met for all equations. In the following, the proposed algorithm is first validated with consideration of the benchmark problem of the sedimenting sphere, and the acquired results are compared to the previously reported results in the literature. Subsequently, this method is used to simulate electrified Newtonian and viscoelastic jets and an in-depth analysis of the results is given.

### A. Benchmark Problem of the Sedimenting Sphere

To validate the viscoelastic constitutive equations and their implementation, the benchmark problem of a sedimenting sphere is modeled. Many researchers have considered this problem[30, 31]; nevertheless, the Knechtges study[31] is chosen as our main reference due to the similar viscoelastic constitutive equations and the LCM reformulation used in this work. This benchmark problem can easily demonstrate the ability of the LCM to solve sophisticated viscoelastic problems because viscoelastic stresses are resolved in a purely extensional flow in the wake of the sphere while the flow is subjected to a contraction–expansion cross section. The normal component of the conformation tensor in the flow direction is chosen as the main validation factor. The computational domain, shown in Fig. 1, and the physical properties of the fluid, listed in Table I, are identical to the values used in the previous study[31].

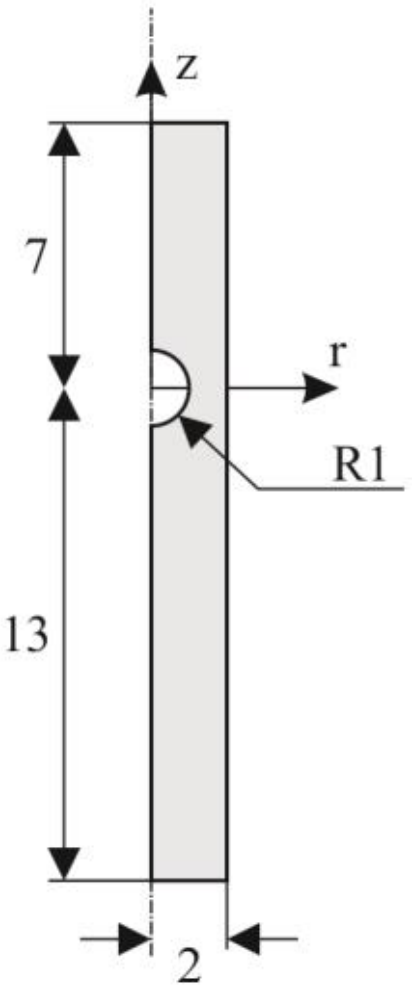

Figure 1. The computational domain utilized in the benchmark problem of the sedimenting sphere. All dimensions are given in millimeters.

Table I. Physical properties of the fluid used in the benchmark problem of the sedimenting sphere.

| Density $\left({kg}/{m^3}\right)$ | Solvent Viscosity $(Pa.s)$ | Polymeric Viscosity $(Pa.s)$ |
|---|---|---|
| 1000 | 0.5 | 0.5 |

In this subsection, inertia terms are neglected and the Stokes flow equations are solved. Moreover, the no-slip boundary condition is applied to the wall of the sphere, and the gravity effect is neglected. Uniform velocity and the constant-pressure boundary conditions are applied at the inlet and outlet of the domain, respectively. It should be noted that, due to the long inlet length, flow in the channel becomes fully developed before reaching the sphere. In addition, viscoelastic stresses are assumed to be zero at the inlet, and their flux is set to zero at the symmetry axis. Except for those indicating convection, the terms in LCM equations are treated explicitly using the previous iteration values. A grid study is performed using three different triangular meshes with their properties listed in Table II.

Table II. Different properties of the triangular meshes utilized in the benchmark problem of the sedimenting sphere.

| Mesh | Number of Domain Elements | Number of Boundary Elements |
|---|---|---|
| M1 | 6644 | 455 |
| M2 | 11791 | 537 |
| M3 | 32206 | 701 |

The results are reported when the solution reaches a steady state. The characteristic length and velocity used in the definition of the Wi are the radius of the sphere and the mean value of the fully developed velocity profile, respectively. Notably, the computational time required to attain a steady-state solution rapidly surges as the Wi increases, because the wake requires a considerably longer time to develop. Fig. 2 shows the normal component of the conformation tensor on the symmetry axis in the wake area of the sphere for the three mesh sizes listed in Table II and three different Wis. The results depicted in the figure show good agreement with the results provided by Knechtges[31], which supports our implementation of the LCM. As reported previously, when the Wi is increased beyond 1, mesh convergence is gradually lost due to the large growth of stress in the wake of the sphere and the amplification of slight deviations because of the exponential function in the LCM equations.

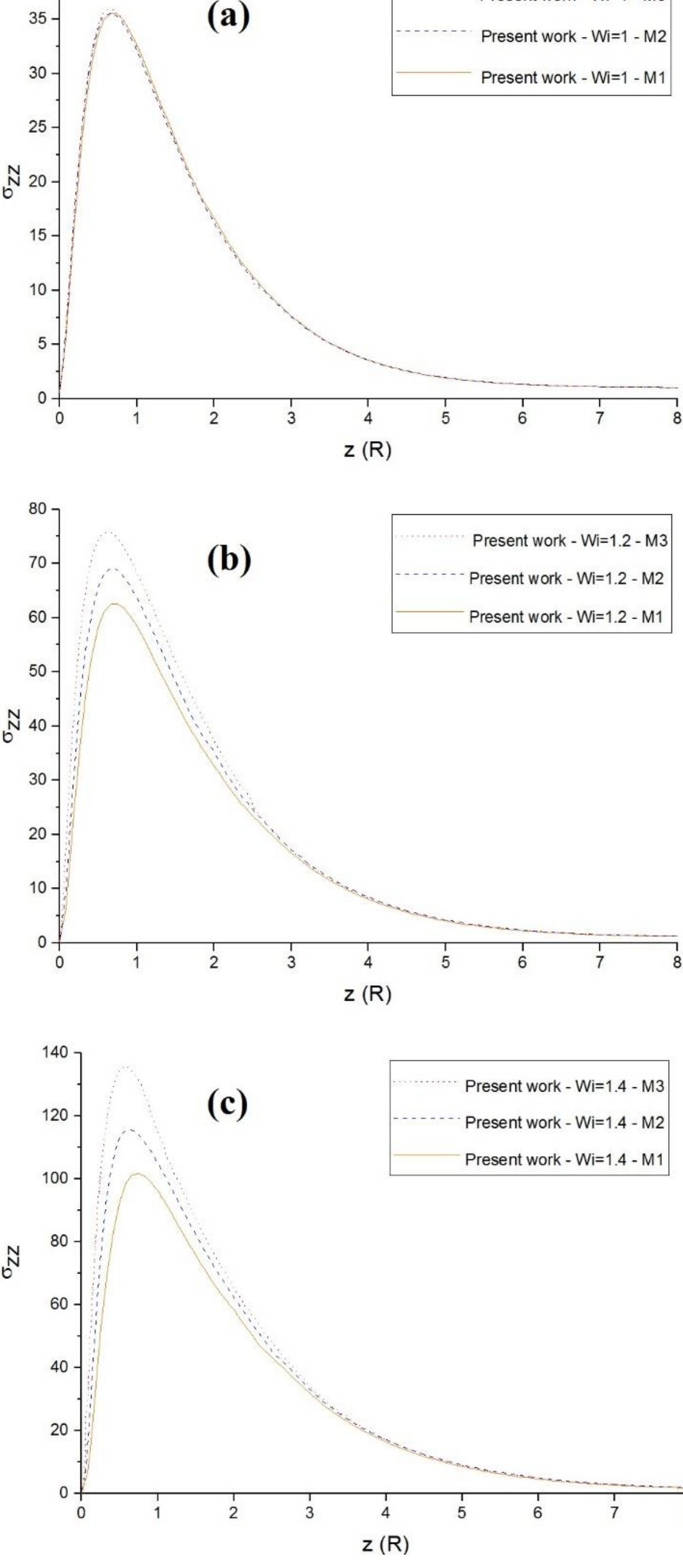

Figure 2. The normal component of conformation tensor in the flow direction on the symmetry axis in the sphere wake area obtained from current simulations for three different mesh sizes and Weissenberg numbers: (a) Wi = 1, (b) Wi = 1.2, and (c) Wi = 1.4.

Furthermore, the contours of the normal component of the conformation tensor in the flow direction are plotted in Fig. 3 for Wi = 1.4 and a mesh size of M2, where the elongation of the viscoelastic stresses along the flow direction is visible. The elongated stresses are initially formed in the wake of the sphere and grow gradually as the wake grows. These contours follow the pattern of pressure modifications behind the sphere due to the pulling of the wake and the action of drag forces in the shear layer[30].

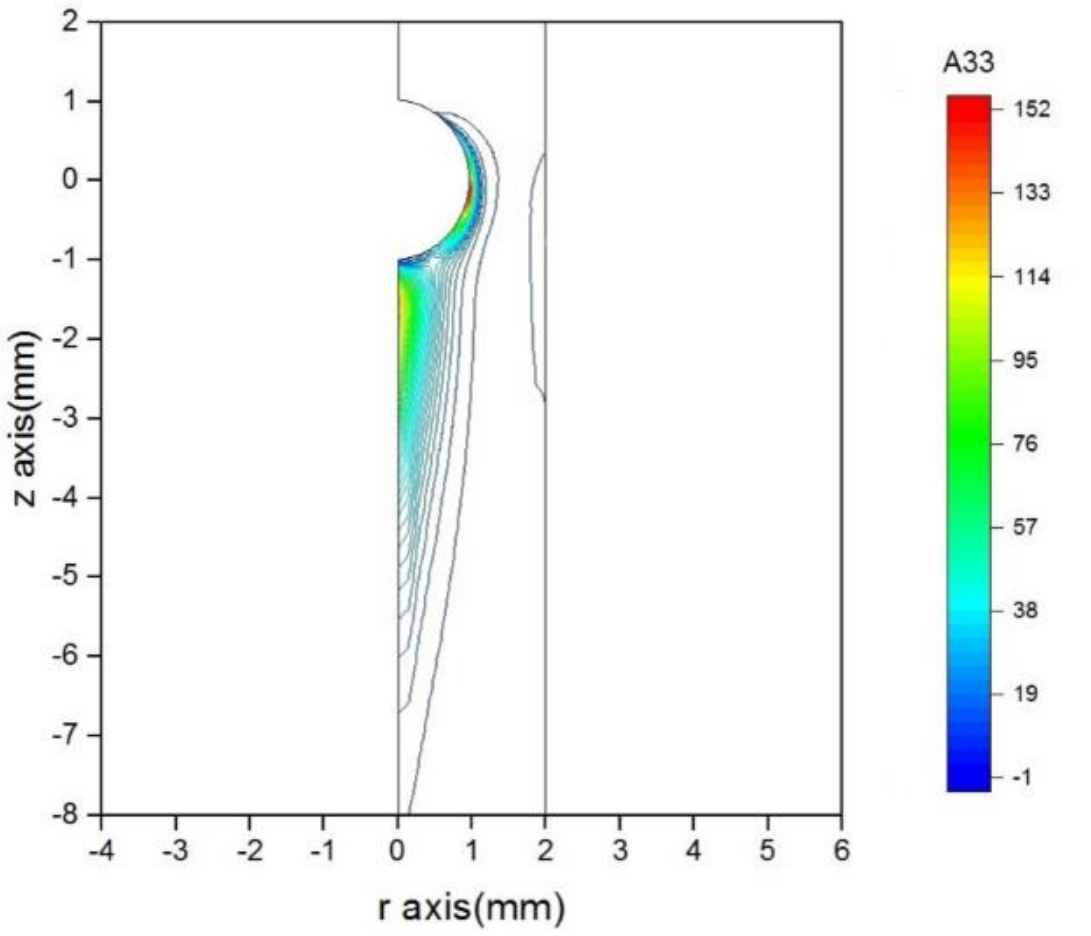

Figure 3. Contours of the normal component of conformation tensor for Wi = 1.4 and a mesh size of M2 in the benchmark problem of the sedimenting sphere.

Finally, to find the stability threshold of this benchmark problem, several numerical simulations are conducted for an increasing sequence of Wis (Fig. 4). This limit was previously reported by Knechtges[31] to be Wi = 1.4; nevertheless, our results show that the Wi can be increased up to 2.6 without the appearance of any oscillations in the solution. This increase in the stability threshold arises from the utilization of the LCM reformulation in addition to the Petrov-Galerkin finite-element method and is a vital factor for the success of electrospray simulation, where the Wi is inherently large due to the small characteristic length.

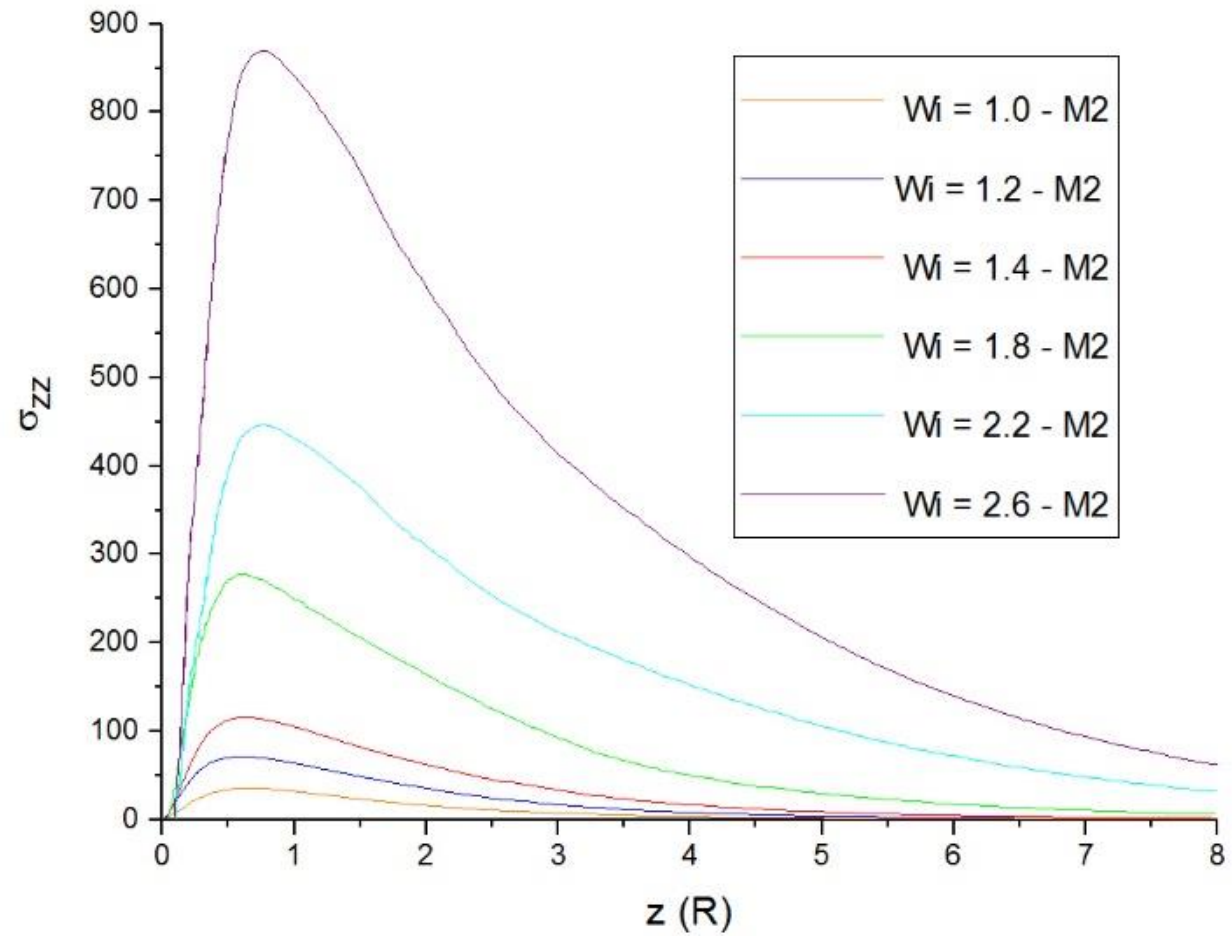

Figure 4. The normal component of conformation tensor on the symmetry axis in the wake area of the sphere obtained from current simulations for a mesh size of M2 and different Weissenberg numbers in the benchmark problem of the sedimenting sphere.

## B. DI Water Jet Simulation

First, we examine our model by simulating an electrified DI water jet, meaning that viscoelastic effects are temporarily neglected. The domain geometry consists of a nozzle positioned on top of the computational domain, as shown in Fig. 5. The dimensions of the nozzle are obtained from our previous experimental work. The boundary conditions used for solving the electrostatic equations are as follows: a constant potential of 11 kV is applied to the walls of the nozzle, while the lower boundary is set as the substrate. For other boundaries, normal electric displacement is set to zero. To reduce the computational cost, a very fine mesh is used in a narrow region close to the symmetry axis where the jet is developing, and the remainder of the domain is covered by a coarser mesh, with a smooth growth factor. Three different mesh sizes were utilized in the central area of the domain to examine the mesh dependency of the results. The detailed information concerning these grids is listed in Table III. The mesh study is done for the viscoelastic jet, and its results are brought in the next subsection; however, since M2 and M3 mesh sizes yield almost the same results, the DI water jet is simulated by the M2 mesh size. The electrostatic equations are discretized with the Galerkin finite element method using quadratic shape functions. Similarly, the coupling between the electrostatic and other equations is accomplished using an iterative, segregated method.

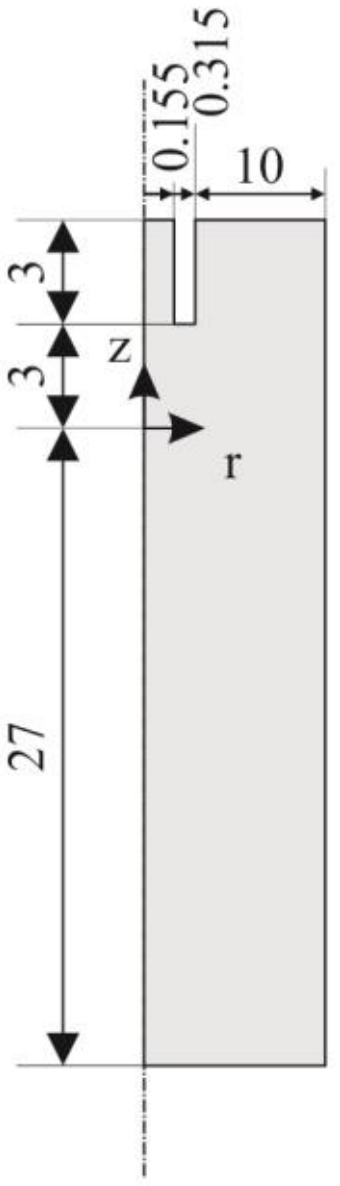

Figure 5. Domain geometry of DI water and viscoelastic solution jet simulations. All dimensions are given in millimeters.

Table III. Various properties of the triangular meshes utilized in DI water and viscoelastic solution jet simulations.

| Mesh | Maximum Element Size (mm) | Number of Domain Elements |
|---|---|---|
| Coarse | 0.428 | 3956 |
| Fine 1 (M1) | 0.006 | 27407 |
| Fine 2 (M2) | 0.004 | 46971 |
| Fine 3 (M3) | 0.002 | 152260 |

The simulated jet profile for a 108 mL/h flow rate, an 11 kV applied voltage, and a mesh size of M2 is shown in Fig. 6. In the simulation results, it is demonstrated that miniscule droplets are detached from the tip of DI water jet, in agreement with the results acquired by Narvaez Munoz[56] and the ramifying behavior observed at the tip of DI water jet in our experimental tests. However, the breakup process is not precisely the same as the experimental results, due to the axisymmetric limitations imposed on the problem. By utilizing axisymmetric coordinates, azimuthal components are set to zero; therefore, the movement of instability waves, ramifications and branches in the third dimension cannot be simulated. In addition, the distance at which drop detachment occurs in the simulation is approximately eight outer diameters of the nozzle, while the ramifying behavior in experimental tests begins after about 12 diameters. In experiments, the main body of the jet has the shape of an irregular cone while in axisymmetric coordinates it takes an almost cylindrical shape. Having a higher surface area than its corresponding conical jet, the cylindrical jet accommodates more charges on its surface, causing the breakup of droplets to happen at a smaller length from the tip of the nozzle.

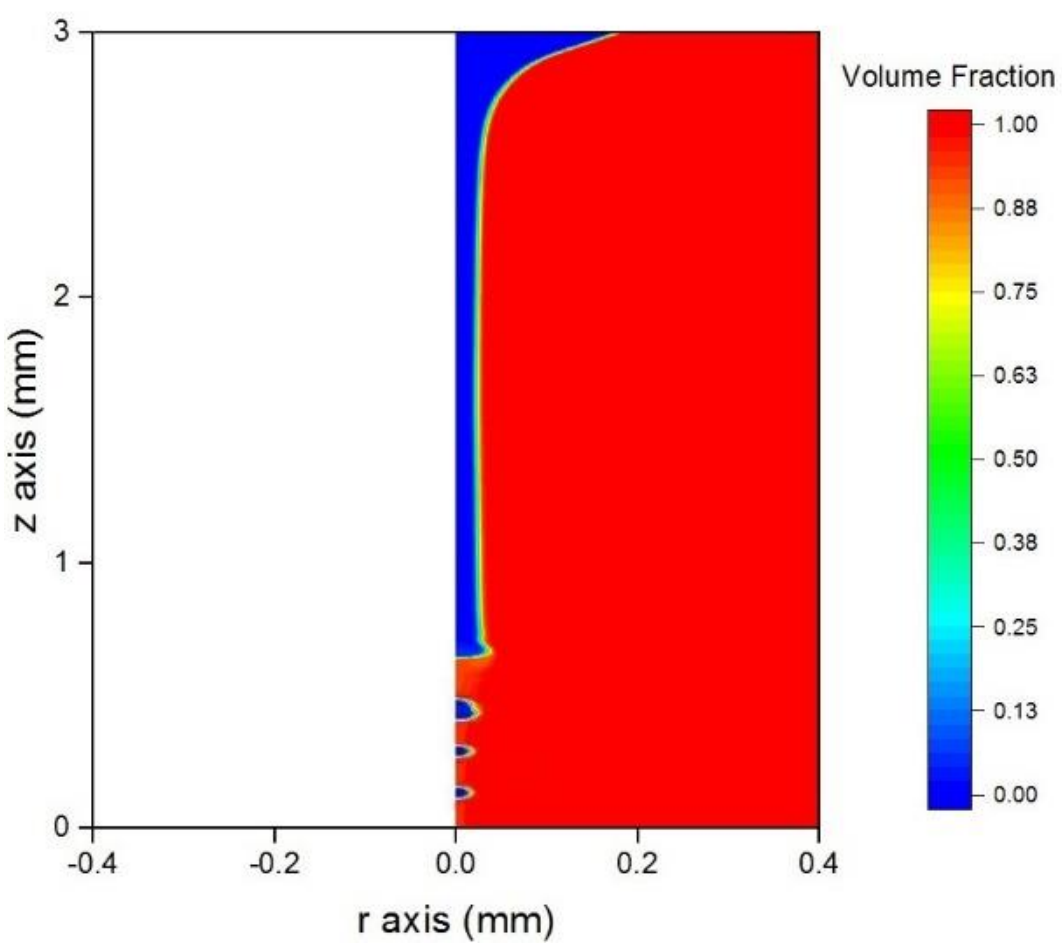

Figure 6. Results of DI water jet simulation for a 108 mL/h flow rate (We = 1.382), an 11 kV applied voltage ($Ca_E$ = 2.69), and a mesh size of M2.

In Fig. 7, a comparison is made between the simulated jet profile and its corresponding profile obtained from image-processing data of our experimental tests. By using image-processing codes on the snapshots acquired by high-speed photography, the noise on the jet surface and its surroundings is eliminated, and the error equals one pixel, which is also taken into account in the experimental data shown in the figure. It is clear from the figure that the simulation results predict a higher cone angle, and the tip of the nozzle is more wetted in the experiment. On the other hand, the asymptotic thickness of the jet is reached at a smaller length from the nozzle in the numerical results. The observed discrepancy between the two jet profiles may be rooted in several sources, including the deviation of the electric field in the experiment from the ideal field used in the computation, due to the presence of a supporting frame and other measurement devices near the nozzle. Other major sources of error include the error in voltage measurement (0.1 kV), the error in flow rate measurement (1 mL/h), and the disparity between charge distribution patterns on the jet surface in the simulation and experiments, as described earlier.

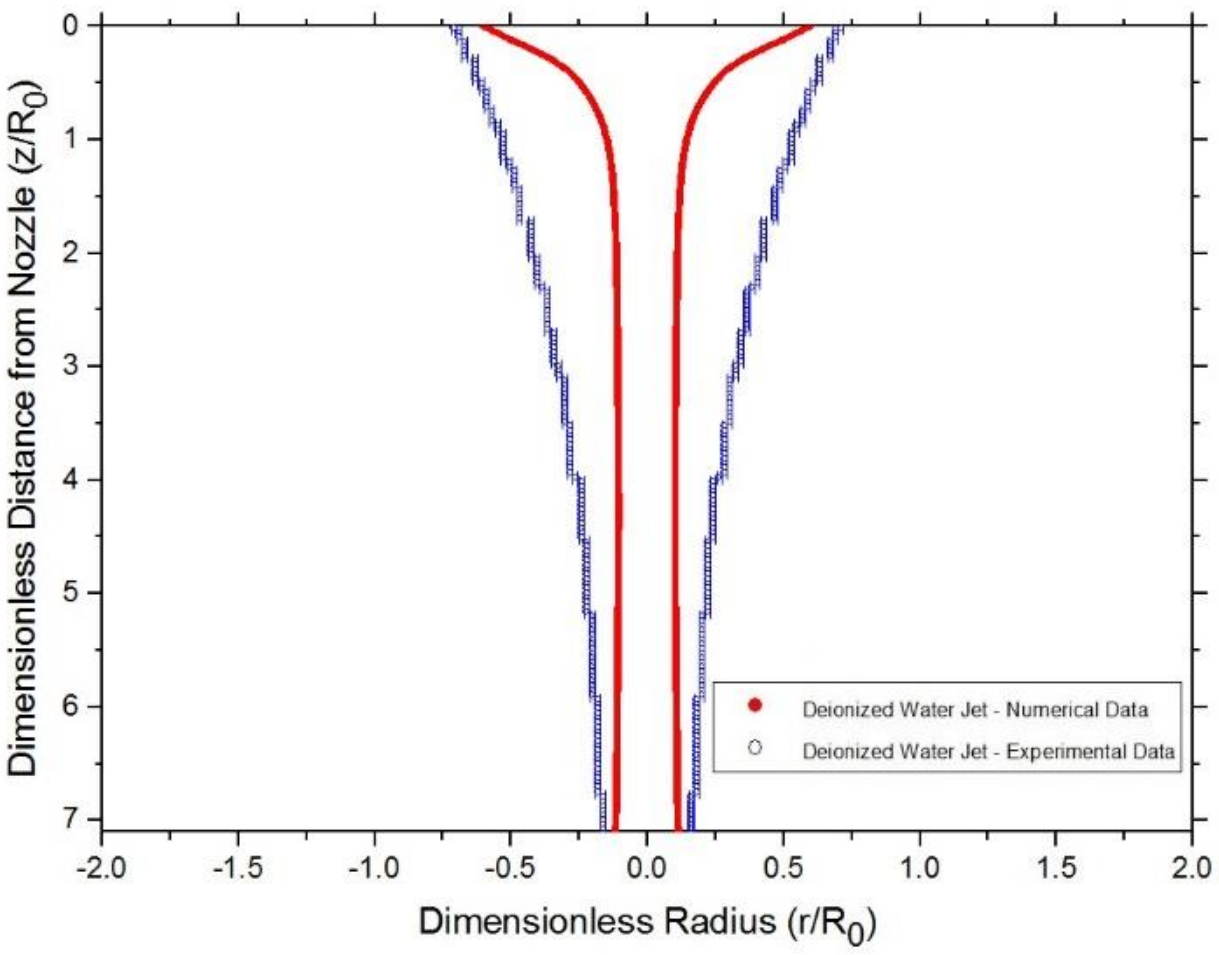

Figure 7. Comparison between simulated DI water jet profile and its corresponding profile acquired from image-processing data. The jet profiles are plotted until the point where droplet breakup occurs in the simulation. For visualization purposes, the profiles are mirrored with respect to the central symmetry axis.

### C. Viscoelastic Solution Jet Simulation

The computational domain utilized for the viscoelastic electrified-jet simulations is the same as the one used in the previous subsection for DI water. The boundary conditions used when solving the LCM and electrostatic equations are identical to the boundary conditions of the sedimenting sphere and DI water jet simulations, respectively. The viscoelastic stress equations are discretized using the Petrov–Galerkin finite element method and coupled to the previous equations with an iterative segregated approach, as described earlier. The numerical simulations of viscoelastic jets were done with various grid sizes, where the mesh size in the center of the computational domain was changed; however, mesh sizes bigger than 0.006 mm could not properly capture the fine viscoelastic jet and its respective interface in the asymptotic region. The mesh study for the viscoelastic jet was done with three different triangular meshes, as listed in Table III, and the acquired results are plotted in Fig. 8. It can be seen that our solutions are independent of mesh size; thus, to reduce the computational costs, every jet simulation in this work was done with M2 mesh size.

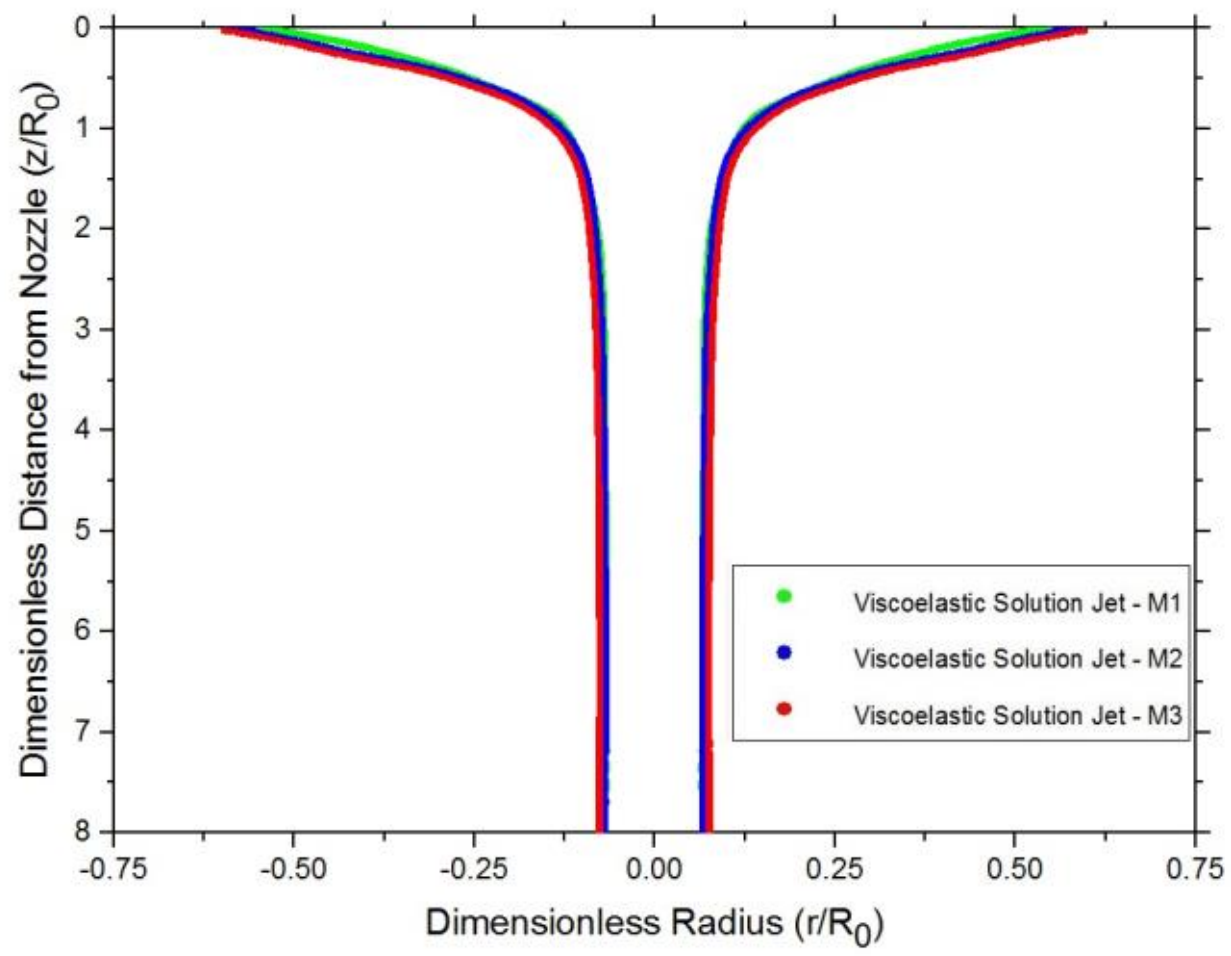

Figure 8. The obtained viscoelastic jet profiles from three different mesh sizes for a 100 ppm PAA solution (Wi = 690.2), a 108 mL/h flow rate (We = 1.387), and an 11.4 kV applied voltage ($Ca_E$ = 2.88). For visualization purposes, the profiles are mirrored with respect to the central symmetry axis.

The jet profile for a 100 ppm PAA solution, a 108 mL/h flow rate, an 11.4 kV applied voltage, and a mesh size of M2 is demonstrated in Fig. 9. Major differences are observed between DI water and viscoelastic solution jet profiles. By contrast to DI water, the viscoelastic jet is stable, and no breakup is seen in the results. Similar behavior was also observed in our experimental results. In agreement with the experimental data, for the same operating parameters, the viscoelastic jet is markedly thinner than the DI water jet. These changes in the jet behavior can be attributed to the elastic intermolecular forces in polymers and the extension of viscoelastic stresses in the main body of the jet when it is deformed against tangential electric stresses. The elongation of stresses transforms the initially coiled configuration of polymer molecules to the stretched conformation of these molecules in the jet structure. This stretching in polymer networks increases the extensional viscosity, resulting in the more robust elastic forces in polymer chains which in turn prevents drop shedding from the tip of the jet.

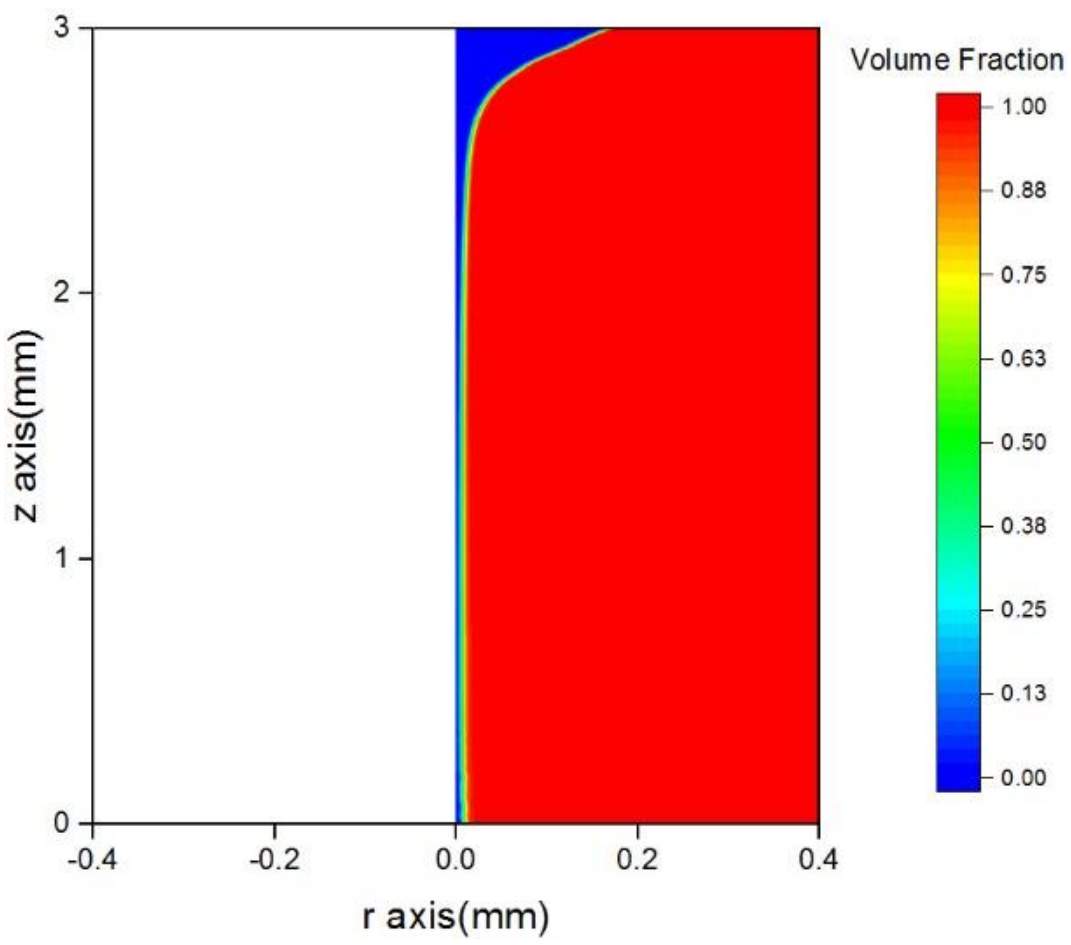

Figure 9. Results of viscoelastic solution jet simulation for a 100 ppm PAA solution (Wi = 690.2), a 108 mL/h flow rate (We = 1.387), an 11.4 kV applied voltage ($Ca_E$ = 2.88), and a mesh size of M2.

In Fig. 10, the space charge density contours are shown in the body of the jet. It can be deduced from the accumulation of the space charge density contours at the extremely thin interface that our leaky dielectric model successfully retains the electric charges close to the interface. Fig. 11 depicts the effects of Wi on the simulated viscoelastic jet profile. As can be seen in the figure, when Wi is increased, the asymptotic profile of the jet is reached at a smaller length from the nozzle, while the final thickness of the jet is slightly reduced. As noted previously in our experimental results, the observed alterations in the jet profile can be attributed to the vigorous stretching of viscoelastic stresses near the nozzle, which is suppressed by the electrostatic forces further downstream. This suppression is amplified as the jet moves away from the nozzle, and the asymptotic thickness of different jets converge, as is evident in Fig. 11.

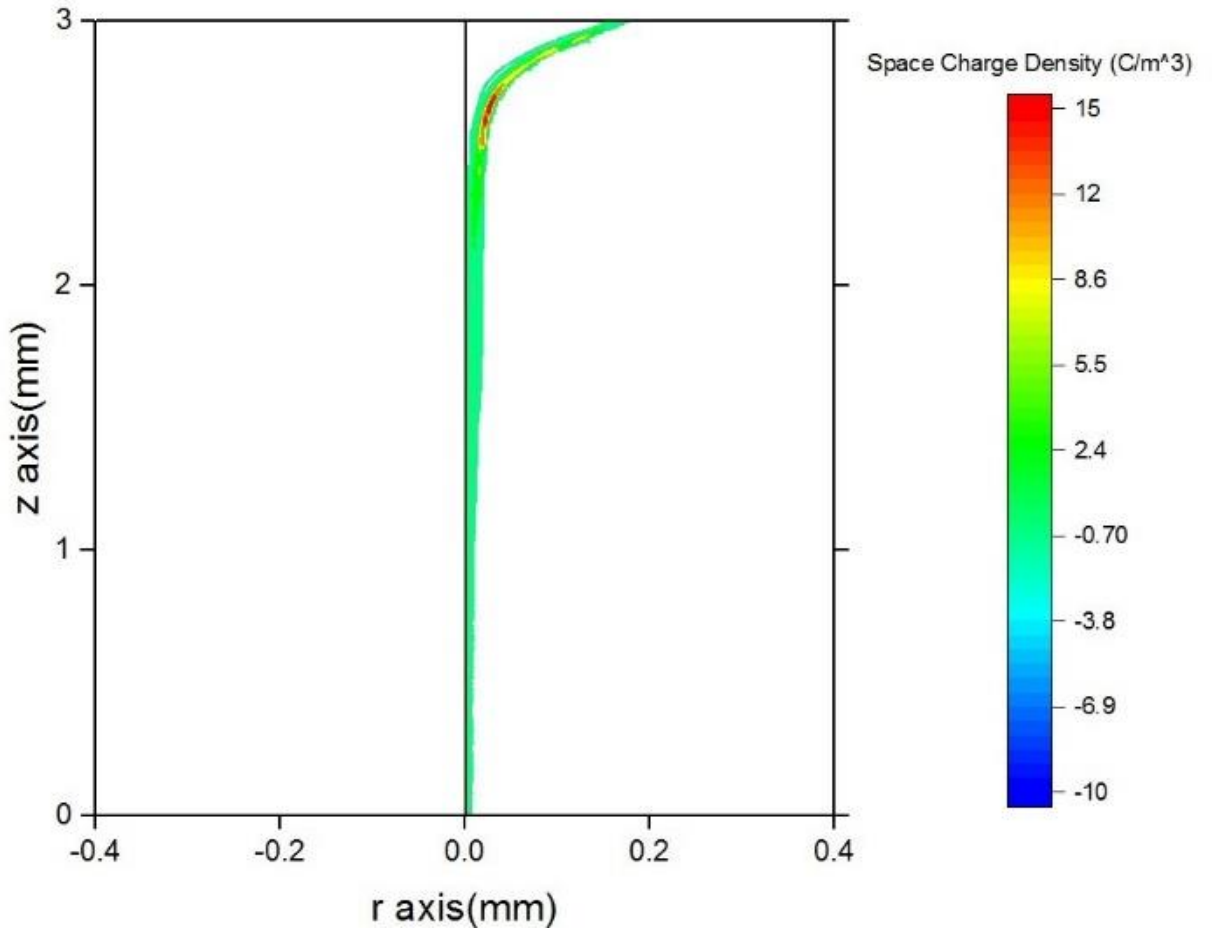

Figure 10. Space charge density contours plotted in the body of the simulated viscoelastic solution jet for a 100 ppm PAA solution (Wi = 690.2), 108 mL/h flow rate (We = 1.387), an 11.4 kV applied voltage ($Ca_E$ = 2.88), and a mesh size of M2.

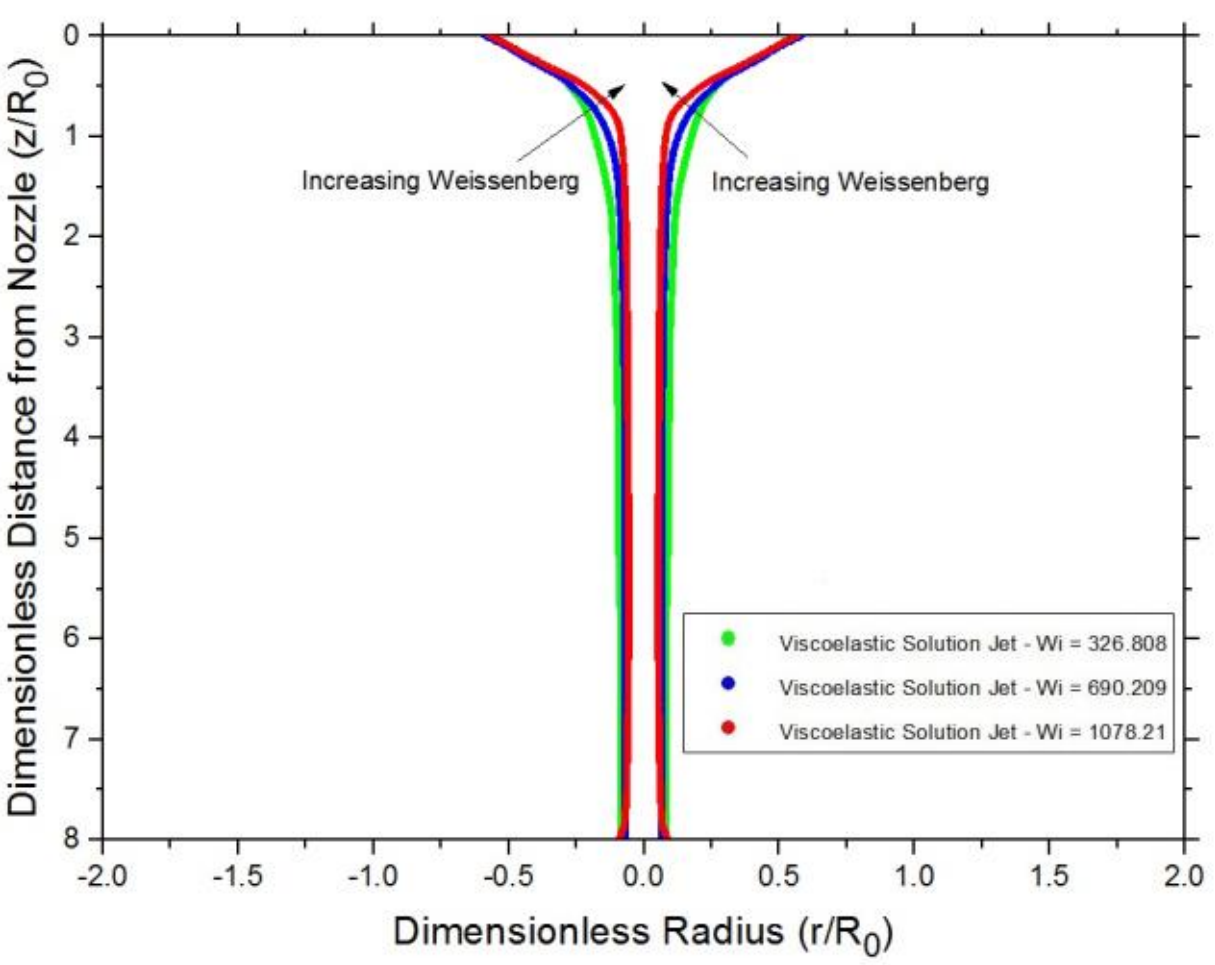

Figure 11. Effects of Wi on the simulated viscoelastic jet profile for a 108 mL/h flow rate (We is not constant due to the variable surface tension between different concentrations) and an 11.4 kV applied voltage ($Ca_E$ = 2.88), and a mesh size of M2. For visualization purposes, the profiles are mirrored with respect to the central symmetry axis.

Fig. 12 compares the simulated results to the experimental data obtained from image processing of high-speed photography snapshots with operating parameters of a 100 ppm PAA solution, a 108 mL/h flow rate, an 11.4 kV applied voltage, and a mesh size of M2. The image-processing error is also taken into account for the experimental data. The discrepancy observed between the two jet profiles is similar to the results for DI water, and its respective reasons are elucidated above.

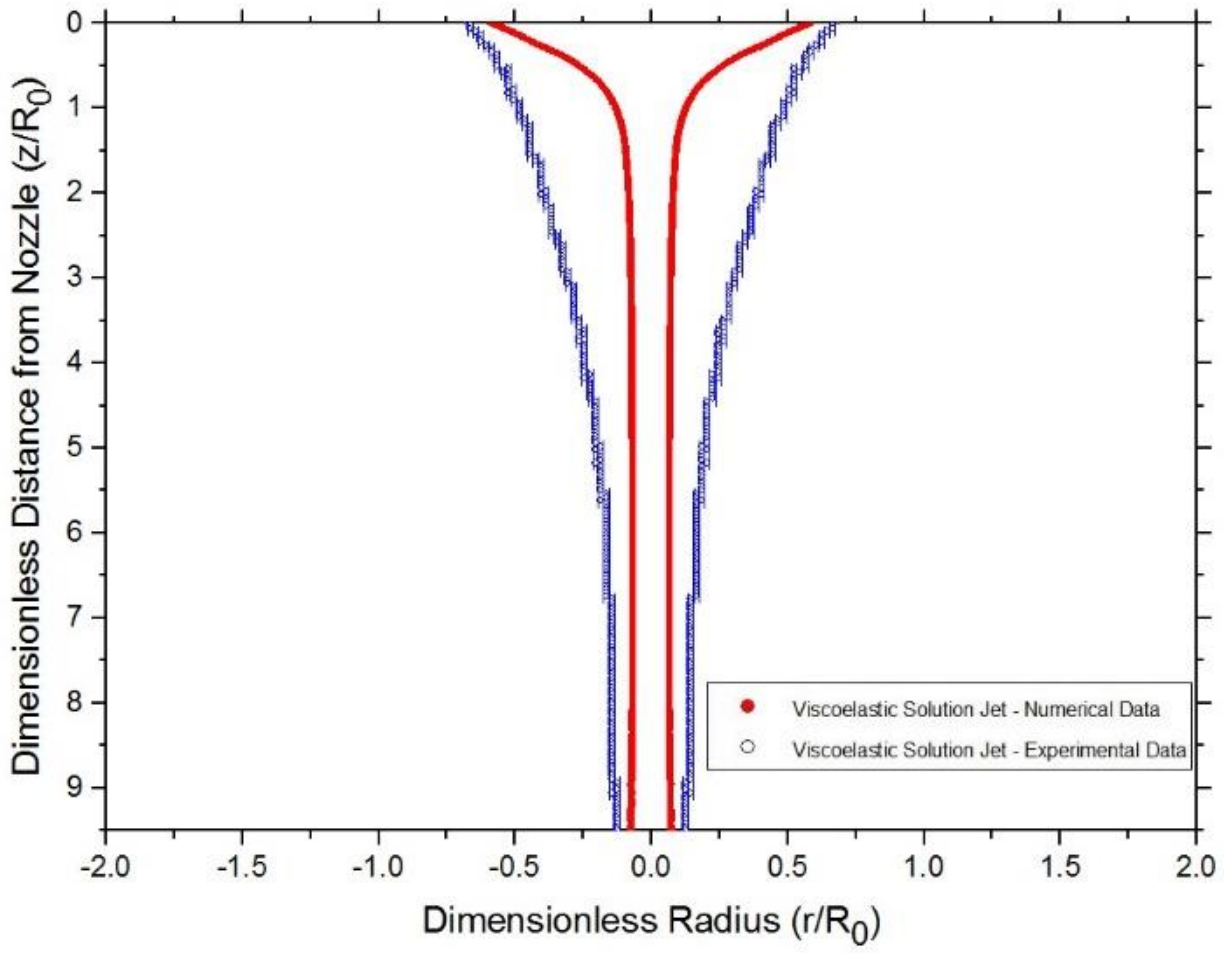

Figure 12. Comparison between simulated viscoelastic solution jet profile and its corresponding profile obtained from image-

processing data. For visualization purposes, the profiles are mirrored with respect to the central symmetry axis.

## IV. Conclusion

In this research, the effect of viscoelastic properties on the jet mode was examined numerically. The alterations in the mechanical behavior of fluid caused by viscoelastic stresses lead to severe changes in jet behavior. It can be concluded from the numerical results that the stabilization limit for the benchmark problem of the sedimenting sphere has increased in our work even though the solution lost mesh convergence when Wi rose beyond 1. Next, the DI water and viscoelastic solution jet profiles were compared with their corresponding experimental profiles. The simulation of viscoelastic jets in an increasing sequence of Wis indicated that when solution concentration surges, the asymptotic profile of the jet is reached at a smaller length from the nozzle, and the final thickness of the jet is slightly reduced. Overall, the effects of viscoelasticity on the simulated jet profiles closely resembled our previously reported experimental results.

## Appendix A

Here, the equations associated with the decomposition of the velocity gradient transpose, together with the relationships used for the computation of eigenvectors and eigenvalues of the log-conformation tensor, are discussed in detail. If $R$ and its transpose are applied to all terms of Eq. (13), and Eqs. (17) to (20) are substituted into Eq. (13), the following relationships are obtained:

$$\omega_{ij} = \begin{cases} 0 & if \quad i = j \\ \dfrac{\lambda_j q_{ij} + \lambda_i q_{ji}}{\lambda_j - \lambda_i} & if \quad i < j \\ -\omega_{ji} & if \quad j < i \end{cases} \tag{A.1}$$

$$n_{ij} = \begin{cases} 0 & if \quad i = j \\ \dfrac{q_{ij} + q_{ji}}{\frac{1}{\lambda_i} - \frac{1}{\lambda_j}} & if \quad i < j \\ -n_{ji} & if \quad j < i \end{cases} \tag{A.2}$$

$$b_{ij} = \begin{cases} q_{ii} & if \quad i = j \\ 0 & if \quad i \neq j \end{cases} \tag{A.3}$$

Using these equations, every term in the decomposition of the velocity gradient transpose can be determined. Additionally, the eigenvalues of the log-conformation tensor are computed with the following equations:

$$\lambda_1 = \begin{cases} \dfrac{\psi_{11} + \psi_{33} + \sqrt{(\psi_{11} - \psi_{33})^2 + (2\psi_{13})^2}}{2} & if\ \psi_{13} \neq 0 \\ \psi_{11} & if\ |\psi_{13}| \leq \varepsilon \end{cases}$$

(A.4)

$$\lambda_2 = \psi_{22} \tag{A.5}$$

$$\lambda_3 = \begin{cases} \dfrac{\psi_{11} + \psi_{33} - \sqrt{(\psi_{11} - \psi_{33})^2 + (2\psi_{13})^2}}{2} & if\ \psi_{13} \neq 0 \\ \psi_{33} & if\ |\psi_{13}| \leq \varepsilon \end{cases}$$

(A.6)

The eigenvectors of the conformation tensor create tensor $R$, which can be computed by:

$$R = \begin{cases} \begin{pmatrix} \frac{1}{A_1} & 0 & \frac{1}{A_3} \\ 0 & 1 & 0 \\ \frac{-\psi_{13}}{(\psi_{33} - \lambda_1) A_1} & 0 & \frac{-\psi_{13}}{(\psi_{33} - \lambda_3) A_3} \end{pmatrix} & if\ \psi_{13} \neq 0 \\ I & if\ |\psi_{13}| \leq \varepsilon \end{cases}$$

(A.7)

where $A_1$ and $A_3$ are defined by the following equations:

$$A_1 = \sqrt{1 + \left(\frac{\psi_{13}}{\psi_{33} - \lambda_1}\right)^2} \tag{A.8}$$

$$A_3 = \sqrt{1 + \left(\frac{\psi_{13}}{\psi_{33} - \lambda_3}\right)^2} \tag{A.9}$$

In Eqs. (A.4) to (A.7), 1e$^{-12}$ is considered for $\varepsilon$ to avoid division by zero. The following equations are the expanded form of Eq. (11) in axisymmetric coordinates:

$$F_{p,r} = \frac{\partial \tau_{11}}{\partial r} + \frac{\partial \tau_{13}}{\partial z} + \frac{\tau_{11}}{r} - \frac{\tau_{22}}{r} \tag{A.10}$$

$$F_{p,z} = \frac{\partial \tau_{13}}{\partial r} + \frac{\partial \tau_{33}}{\partial z} + \frac{\tau_{13}}{r} \tag{A.11}$$